# Edge-Mediated Skyrmion Chain and Its Collective Dynamics in a Confined Geometry

*Haifeng Du,[1,§] Renchao Che,[2,*,§] Lingyao Kong,[3] Xuebing Zhao,[2] Chiming Jin,[1] Chao Wang,[2] Jiyong Yang,[1] Wei Ning,[1] Runwei Li[4] Changqing jin,[5] Xianhui Chen,[1,6] Jiadong Zang,[7,8] Yuheng Zhang,[1,9] Mingliang Tian,[1,9]\**

[1] High Magnetic Field Laboratory, Chinese Academy of Science (CAS), Hefei, Anhui Province 230031, China

[2] Laboratory of Advanced Materials, Department of Materials Science, Collaborative Innovation Center of Chemistry for Energy Materials, Fudan University, Shanghai 200438, China

[3] Institute of Fluid Physics, China Academy of Engineering Physics, Mianyang, Sichuan Province 621900, China

[4] Key Laboratory of Magnetic Materials and Devices, Ningbo Institute of Material Technology and Engineering, Chinese Academy of Sciences, Ningbo, Zhejiang Province 315201, China

[5] Beijing National Laboratory for Condensed Matter Physics and Institute of Physics, Chinese Academy of Sciences, Beijing 100190, China,

[6] Hefei National Laboratory for Physical Science at Microscale and Department of Physics, University of Science and Technology of China, Hefei, Anhui Province 230026, China

[7] Department of Physics and Materials Science Program, University of New Hampshire, Durham, New Hampshire 03824, USA

[8] Institute for Quantum Matter and Department of Physics and Astronomy, Johns Hopkins University, Baltimore, Maryland 21218, USA

[9] Collaborative Innovation Center of Advanced Microstructures, Nanjing University, Jiangsu Province 210093, China

**Corresponding authors***: rcche@fudan.edu.cn (C.R.); tianml@hmfl.ac.cn (M.T.).*

*§The authors contribute equally to this work*

**The emergence of a topologically nontrivial vortex-like magnetic structure, the magnetic skyrmion, has launched new concepts for memory devices. There, extensive studies have theoretically demonstrated the ability to encode information bits by using a chain of skyrmions in one-dimensional nanostripes. Here, we report the first experimental observation of the skyrmion chain in FeGe nanostripes by using high resolution Lorentz transmission electron microscopy. Under an applied field normal to the nanostripes plane, we observe that the helical ground states with distorted edge spins would evolves into individual skyrmions, which assemble in the form of chain at low field and move collectively into the center of nanostripes at elevated field. Such skyrmion chain survives even as the width of nanostripe is much larger than the single skyrmion size. These discovery demonstrates new way of skyrmion formation through the edge effect, and might, in the long term, shed light on the applications.**

**Introduction**

Since its discovery in 2009, magnetic skyrmion has attracted lots of attention in condensed matter physics due to its rich physics as well as the potential applications in spintronic devices[1-4]. It is a vortex-like spin textures, in which local magnetic moments point in a concerted manner to wrap a sphere[5,6]. The peculiar twists of the spins within the skyrmion makes it possess nontrivial topology, which brings an emergent electromagnetic field on electrons passing through, yielding unconventional spin-electronic phenomena[7] such as the ultra-low threshold current density for skyrmion motion[8]. This property, together with their tunable small size and topological stability, offers the skyrmion a promising candidate to build next generation's high density and low energy-consumption memory devices. A representative example is the skyrmion-based race-track memory, where the nanostripe is encoded via control of individual skyrmions using spin polarized current pulses [9-11]. Recently, many theoretical studies have demonstrated the possibility of hosting and manipulating individual skyrmions in such confined geometry[12-14]. However, such important skyrmions arrangement in the nanostripe is so far still invisible experimentally[15].

The capacity to image the spin configurations in nanostructured helical magnets can provide valuable insight into this issue. High resolution Lorentz transmission electron microscopy (TEM), which uses the Fresnel method (See the Supplementary Note 1) to obtain the nanometer-scale configuration of magnetic structures, has achieved great success in identifying skyrmion crystal in helical magnetic thin films[2,16]. In principle, this technique is able to be adapted to observe the magnetic state in ultra-narrow stripes. However, the sharp interface between sample and vacuum would lead to strong Fresnel fringes and give rise to artificial magnetic contrast around the edge. As a result, the real spin configurations around the edge are severely smeared out in such small-size samples[15-17]. Another challenge is the difficulty in obtaining nanostructured helical magnets with the desirable geometry. Recent advance has synthesized high quality MnSi nanowires by chemical method, but the complex crystal structure in the nanowires, such as the parallelogram cross-section and twins boundaries, make them challenge to obtain skyrmion chain under the applied magnetic field perpendicular to the nanowire though individual skyrmion cluster states has been identified from the magnetoresistance measurements when the magnetic field is applied along the long axis of the nanowire[18,19]. Here, we fabricated the nanostripes with the desirable geometries by a top-down method. By further coating the nanostripe with the amorphous PtCx at the edge, we significantly reduce the interfacial Fresnel fringes in Lorentz TEM images. Eventually, a single skyrmion chain (SSC) was unambiguously observed in the nanostripe as its width is comparable to the featured skyrmion size. Further systematic investigation of the magnetization dynamics of nanostripes with various widths unveils an edge-mediated mechanism to create skyrmions in confined geometries at low temperature.

**The magnetization process in 130 nm FeGe nanostripe**

We choose B20 compound FeGe to investigate the effect of confined geometry on the formation and stability of skyrmions. FeGe is a typical helical magnet harboring a skyrmion phase near room temperature. Bulk FeGe samples exhibit a paramagnet to helimagnet transition at a Curie temperature $T_c$~280 K under zero magnetic field[20]. The spin helix possesses a long range wavelength, $\lambda$ ~70 nm, and a fixed wave-vector, $Q$, along the high symmetry crystal axis. If a finite magnetic field, $B$, is applied at the temperature $T<T_c$, it becomes energetically favorable to form a conical phase with $Q\|B$. At an even higher field, the conical phase transfers into ferromagnetic spin alignment. Both helical and conical states belong to single-twist magnetic structure since the rotation of their magnetization only in one direction fixed by the wave-vector $Q$. In contrast, in the skyrmion phase the magnetization rotates in two directions, forming double-twist modulations, which occupy only a tiny pocket in $T$-$B$ diagram slightly below $T_c$ (see Supplementary Figure 3) in bulk samples. This region of stabilizing skyrmion phase would be, to some extent, expanded in the $T$-$B$ space when the sample becomes extremely thin[2,16]. The evolution of spin configurations with varied $B$ and $T$ in FeGe reflects a common behavior in helical magnets[6]

To study the effect of geometric confinement on the formation and stability of skyrmions, a series of FeGe nanostripes with various widths were fabricated from the bulk by a complex process (see supplementary note 2). Figure 1a shows a typical TEM image of the nanostripe with a width of $w$ ~ 130 nm, which is comparable to the featured skyrmion lattice constant, $a_{sk}$ ($2\lambda/\sqrt{3}$ ~ 81 nm). The surrounding gray part is the amorphous PtCx, which can significantly reduce the effect of Fresnel fringes on the real magnetic structure around the edge and enable us to observe directly the magnetic structure in the ultra-narrow stripes (Supplementary Figure 5). An external magnetic field, $B$, is applied perpendicular to the stripe plane with its direction pointing upward (marked as black "⊙"). Figure 1b-h shows the evolution of the spin textures of this nanostripe with increasing $B$ measured at 100 K, where the planar magnetic distributions were obtained by the magnetic transport-of-intensity equation (TIE) analyses[2] of the high-resolution Lorentz TEM data (Supplementary Note 1). The two FeGe/PtCx interfaces, marked by the white dotted lines, would induce weak artificial magnetic contrast in the PtCx region due to the Fresnel fringes (Supplementary Note 3). In the following, we will focus the magnetic state in the FeGe parts by marking the boundary with the small black triangular. At low magnetic field, spin helix are observed with a period of 70 nm (Figure 1b), consistent with the period in the bulk sample[16]. But, the wave-vector, $Q$, is almost parallel to the long axis of the nanostripe. With increasing $B$, the magnetization undergoes a series of dynamical processes (Figure 1c and d), and eventually evolves to circular skyrmions at $B \sim 3.3$ kOe (Figure 1e and e1). These skyrmions assemble a perfectly single chain due to the transverse constriction. This single skyrmion chain (SSC) in the interior of the nanostripe is accompanied by a chiral magnetization on the edge, as indicated by the red arrows in Figure 1e1. Spins on the two edges tilt onto the plane, and orient their planar components parallel to the boundary (schematic diagram in Figure 1e2). That is because of an imbalanced torque acting on the edge spin originating from the DM interaction[10,11,21]. Reversal of the sample chirality would flip the edge spin orientations. Higher fields above $B \sim 4.2$ kOe melt the perfect SSC state into isolated skyrmions confined in the nanostripe (Figure 1f and g), and finally polarize all the interior magnetizations at about $B \sim 4.6$ kOe (Figure 1h), where magnetic contrast of the planar components disappears. However the chiral edge magnetization is still persistent.

At elevated temperatures, the SSC still persists. The $T$-$B$ phase diagram of the 130 nm nanostripe (Supplementary Figure 6) explicitly shows a highly stable SSC in this confined geometry, where the SSC survives in the whole temperature region accessible by our Lorentz TEM experiments. This lowest temperature (~ 100 K) is far below the magnetic transition temperature $T_c$ (~ 280 K)[20] and is even much lower than 200 K, the lower bound of the skyrmion phase reported in 2D FeGe plates with the same thickness of $t$ ~ 65 nm to the 130 nm nanostripe[16]. Very likely, the SSC phase extends through the whole low temperature regime in this confined geometry. The comparable size between the nanostripe and single skyrmion allows us to identify that such narrow stripe ($w$ ~130 nm) can only accommodate one skyrmion in the transverse direction by simple geometry analyses. However, the wider extension of temperature for hosting skyrmion phase in the nanostripe is sharp contrast to the above-mentioned low stability of skyrmions in bulks[20] or the thickness-dependent stability in two dimensional films[16]. Obviously, this observation suggests an important role played by geometric confinement on the formation of skyrmions.

To explore the physical origin of this effect, we closely examined the magnetization process. Figure 1b shows the spin helix at $B$~0 Oe. Unlike the locking of wave vector, **Q**, along the high symmetry crystal axis due to the weak crystal anisotropy reported so far [6], we notice that the spin helix in narrow stripes always orients its wave vector parallel to the edge of the nanostripe. Such sample-shape dependent orientation of stripe domains has been addressed in the perpendicular anisotropic magnet[22,23], where lowering the stray field energy gives rise to perpendicular or parallel stripe domains at the edge of the sample, with all other orientations being energy less favorable. The origin might be applied here. In this case, spin helices are terminated at the edge, forming periodically modulated "half-disk" domains (the details on the edge state are depicted in Supplementary Note 3). Magnetic moments in these domains have fixed swirling directions (Figure 1b1 and b2), which are indicated by white (clockwise) and red (counter-clockwise) arrows, respectively. Moments at the disk centre are out of plane with staggered polarity upward (marked as "●") and downward (marked as "×"). Although the in-plane magnetization can be directly obtained by TIE analysis, the polarity is inferred by tracing the evolution of magnetization under high magnetic fields, using the fact that a large enough $B$ aligns spins along the same direction. Owing to the chiral property of helical magnets, the rotation direction and polarity of disks are entangled so that only two types of "half disk" domains can occur. These two types of domains alternately appear along the edge since they are coupled to the periodically modulated helices inside the stripe.

When a weak external magnetic field is applied, the Zeeman energy enforces an increase in the proportion of the upward spins. According to the Lorentz TEM data, the increasing number of upward spins is accommodated by expanding the boundaries of the red "half-disk" domain (the moving directions are marked by small black arrows with white frame) as its polarity in the center is parallel to the magnetic field. Above a threshold value of magnetic field $B \sim 1.8$ kOe, the edge spins coalesce in the red parts, forming a uniform edge state (red straight arrows in Fig. 1c1 and details on Supplementary Figure 7)[11,12,21], while the white parts are lifted away from the edge, forming half skyrmions. i.e. merons[24]. Another meron is similarly formed at the opposite edge, and together they constitute a bimeron (Figure 1c1 and c2), which can be regarded as an elongated skyrmion as it carries the same unit topological charge as a skyrmion. With further increase of magnetic fields up to ~ 870 Oe, the elongated skyrmions are compressed (Figure 1e1 and 1e2), and form a perfect SSC with the topological charge conserved.

**The magnetization process in 396 nm FeGe nanostripe**

This magnetization process in the 130 nm nanostripe provides a strong hint that the creation of skyrmions in confined geometries is closely related to the presence of "half-disk" domains, originated from spin helices propagating along the edge. To test this idea, we systematically investigated the magnetization dynamics of nanostripes with increasing width while fixing the low temperature $T \sim 100$ K. Figure 2a-i. shows a representative evolution of spin textures in a $w \sim 396$ nm nanostripe. The corresponding enlarged images are selectively illustrated in Figure 2a1-e1. Compared with the $w \sim 130$ nm stripe, this 396 nm stripe is nearly 5 times larger than the skyrmion lattice constant ~ 81 nm, and shows a helical state with the co-existence of orthogonal wave vectors **Q** ∥ edge and **Q** ⊥ edge (Figure 2a and a1). Similar to the $w \sim 130$ nm nanostripe, the bottom spin helices with **Q** ∥ edge possess distorted edge state. In contrast, the upper spin helices with **Q** ⊥ edge, being parallel to a uniform edge state, exhibits a perfect helical phase without distortion.

When the magnetic field is turned on, the two orthogonal helices show remarkable differences. At $B \sim 1.6$ kOe (Figure 2b and b1), the helix with **Q** ⊥ edge becomes less visible, while the helix with **Q** ∥ edge begins to deform in the same way as the $w \sim 130$ nm nanostripe. When the magnetic field is further increased to ~ 1.8 kOe, the helix with **Q** ⊥ edge disappears, while the other with **Q** ∥ edge gives rise to compressed bimerons (Figure 2c1), which eventually self-organize to skyrmion chains positioned at the edge at elevated magnetic field (Figure 2d and d1). This observation indicates that skyrmions can only be created from helices with distorted edge state. This scenario works well at low temperatures where the magnetization dynamics are more important than the statistics. In the process, the high mobility of skymions is also illustrated in the confined geometry, where the chain is easy to move under the action of the magnetic fields. This leads to the unfixed positions of skyrmions. For example, a skyrmion chain around the upper right edge at $B \sim 2.2$ kOe in Figure 2d1 merges into the bottom skyrmion chain at $B \sim 3.5$ kOe (see Figure 2e1) without any creation or annihilation of skyrmions. Similar phenomena are also found in 306 nm wide stripe with some certain defects (Supplementary Figure 9f and g) and 550, 1017 nm wide stripes (Supplementary Figure 10). In spite of this high mobility, there is almost a one-to-one correspondence between the number of these helix periods and the number of skyrmions.

Once the skyrmion chain is formed at the edge (Figure 2d) another interesting observation is that the skyrmion chain tends to move collectively into the center of the stripe (Figure 2d-g) with increasing magnetic field, and the number of skyrmions remains almost unchanged over a wide interval of magnetic field strengths (2.2 kOe $< B <$ 4.9 kOe). The fixed number of skyrmions supports the topological stability of skyrmions that they become robust once skyrmions are created[5]. The collective movement of skyrmions can be readily explained by the repulsions between edge spins and skyrmions[25], conversely confirming the well-known edge state. Consequently, a distorted skyrmion chain (DSC) centered in the nanostripe (Figure 2g) is formed. Finally, at even higher magnetic fields with $B > 5.3$ Oe, the topologically stable DSC is melted and skyrmions therein gradually disappear, as a transition to the ferromagnetic spin texture occurs (Figure 2h and i).

This self-organized skyrmion chain at the edge induced by the helices with the distorted edge state and the subsequent collective movements under an action of magnetic fields are found to be a universal phenomenon that is extensively verified at low temperatures in nanostripes of various widths (Supplementary Figure 9 and 10). Moreover, since the lengths of nanostripes are finite in our experiments, the skyrmion chain is also observed at the short edge of wide nanostripes as long as the helices with distorted edge state exist. These experimental results all point to the conclusion that the spin helices with distorted edge state underlie the emergence of skyrmion chains. In other words, we have identified an edge-assisted mechanism to create skyrmion chains in nanostripes instead of the previously reported skyrmion crystal, where thermodynamics plays the key role[2,16]. This self-organized chain in our system is completely distinct from the previously reported preferential formation of skyrmions near the sample edge in 2D plates of FeGe[16]. Those samples, fabricated by traditional argon-ion thinning methods, have an inhomogeneous thickness with the thicker part in the interior and the thinner part at the edges, where skyrmions around the edge are mainly stabilized by the reduced thickness.

**Temperature dependence of skymions arrangement**

To gain a more thorough understanding of skyrmion formation in confined geometries, elevated temperatures were also investigated (Figure 3a-d), where thermal fluctuations would be dominant compared to the magnetization dynamics. It is well understood that a closely packed skyrmion (CPS) crystal is favored by thermal fluctuations in bulk or 2D films. This crystallization of skyrmions is ascribed to inter-skyrmions interactions[6], which is an intrinsic property of skyrmions and should be independent of material details and sample sizes. Therefore the CPS structure is expected in nanostripes when the temperature is high enough. This prediction is clearly supported by Lorentz TEM images of nanostripes, which show that the skyrmions are closely packed (Figure 3a-d and supplementary Figure 11) with the number of skyrmions in the transverse direction proportional to the width of nanostripes.

Based on all the Lorentz TEM data on nanostripes with various widths measured at different temperatures, we constructed a full $T$-$w$ phase diagram of the skyrmions in nanostripes (Figure 3e). While the maximum number, $N_S^{max}$, of skyrmions in CPS that can be accommodated into the defined nanostripe is fixed, the actual skyrmion number, $N_S$, however depends on the temperatures and the applied magnetic field. Here we define $N_S$ at each temperature from the point at which the skyrmion number in the SSC or DSC starts to decrease under further increase of the magnetic field $B$. In other words, $N_S$ is the actual maximum skyrmion number at a defined $T$ under the application of a moderate magnetic field. Clearly, the CPS appears only at high temperatures. At temperatures below about $T < 200$ K, skyrmions in the nanostripe appear in the form of SSC (when $w < 160$ nm) or DSC (when $w > 160$ nm). These results reveal the importance of magnetization dynamics in skyrmion formation at low temperatures.

**Discussion**

We have demonstrated that the edge can be used to control the formation of skyrmions in the confined geometry under the action of the magnetic field. At high temperature, the skyrmions tend to form closely packed structure. This skyrmions arrangement is consistent with previous experimental results on the skyrmion crystal in the corresponding 3D or 2D FeGe compounds[1,2]. In this case, single skyrmion chain can be obtained by reducing the width of the nanostripe to the featured skyrmion size. More importantly, we observed the single skyrmion chain in the nanostripes even as the width of nanostripe is much larger than that of single skyrmion at low temperature. We further identified the skyrmions are originated from the sample-shape dependent orientation of the helical state. The helix with distorted edge state evolves into skyrmions, while those without distorted edge state transfers into ferromagnetic state directly. Taking the topological stability of skyrmions into account, it is commonly concluded that skyrmion cannot easily be destroyed and created from the single-twist magnetic structure including helical and conical phase. By contrast, the confined effect gives rise to the distorted edges states, which show "half-disk" arrangements, indicating their multi-twist modulations. Our results thus indicate the creation of the skyrmions from the multi-twist magnetic state is much easier than that from single-twist magnetic state. These findings offer the fundamental insight into the dynamical mechanism of the formation and evolution of the skyrmions in the confined geometry, and might improve the viability of proposals by use of skyrmions in magnetic nanostripes as the carriers of high-density information through the "edge/or defect-engineering".

In addition, we have further observed the collective motion of the skyrmion chain,

accompanying by a series of adjustment in the chain space and position. This motion is the direct results of skyrmion-skyrmion and skyrmion-edge repulsions. Presence of spins distortion around the edge is fundamental properties in geometry-confined nanostructures. This edge-mediated mechanism together with control of temperature and thickness can provide a powerful approach for tailoring the desired properties of skyrmions in confined geometries.

**Methods Summary**

**Bulk sample preparation**

The polycrystalline B20-type FeGe samples were synthesized with a cubic anvil-type high-pressure apparatus. A mixture of the elemental materials with an atomic ratio of 1:1 was synthesized by electromagnetic induction melting in an argon atmosphere. The alloy was placed into a cylindrical BN capsule and was heat-treated for 1 h at 1073 K under a high pressure of 4 GPa. Structural characterization by X-ray diffraction and susceptibility measurements were consistent with the well-established structure and properties of bulk FeGe.

**Fabrication and characterization of the FeGe nanostripe**

The nanostripes for TEM observation were prepared by lift-out method using a focused ion beam (FIB) and scanning electron microscope (SEM) DualBeam system (Helios Nanolab. 600i, FEI) combined with a Gas Injection System (GIS), and Micromanipulator (Omniprobe 200+, Oxford). The details of the sample fabrication processes and characterization are shown in Supplementary Figure 2. The parameters of fabricated samples with varied width from 130 nm to 1017 nm are summarized in supplementary Table 1. In the main text, we mainly show the data from NS1 ($w \sim 130$ nm) and NS4 ($w \sim 396$ nm). Other data including some typical magnetization processes and the *T-B* magnetic phase diagram are shown in the Supplementary Information.

**Lorentz TEM measurements**

We carried out the magnetic contrast observation by using the Lorentz TEM (JEOL-2100F). In Lorentz TEM mode, the objective lens was turned off and an objective mini-lens under the specimen was employed to induce the normal magnetic field to the thin nanostripe. The observed magnetic contrast can be understood qualitatively in terms of the Lorentz force acting on the moving electrons as they travel through in the magnetic foil. The high-resolution lateral magnetization distribution map was obtained by TIE analyses of the Lorentz TEM images. The details of the Lorentz TEM and the TIE analyses are given in the Supplementary Note 1. Thin-plate thickness was measured using EELS. A double-tilt liquid-nitrogen cooling holder (Gatan, Cryo-Transfer Holder) was used to detect the phase transition below the Curie transition temperature $T_c$. This enables the specimen temperature to be reduced to 100 K with the measured temperature displayed on the cooling holder controller. The magnetic field applied normal to the thin plate was induced by the magnetic objective lens of the TEM.

**Acknowledgments**

This work was supported by the National Key Basic Research of China, under Grant No.

2011CBA00111, 2013CB932901, 2013CB934600; the Natural Science Foundation of China, Grant No. 11274066, No. 11174294, No.11474290, No. 11104281, No. 11374302, No. 11222434, No. U1432251, 51172047, 51102050, U1330118; the Anhui Provincial Nature Science Foundation of China, Grant No.1308085QA16; US National Science Foundation, No. ECCS-1408168; US Department of Energy, Grant No.DEFG02-08ER46544; and the Theoretical Interdisciplinary Physics and Astrophysics Center. The authors thank Peng Chen and Qingqing Liu, Wenming Li, and Weixing Xia for assistance with the sample fabrications and Lorentz TEM observations.

**Author contributions**

H. D., J. Z., R. C., and M. T. conceived and designed the experiment. W. N., J. Y., and C. M. J. fabricated the nanostripes. R. C., X. Z., and C. W. performed the Lorentz TEM observations. The FeGe samples were synthesized by C. M. J., R. L., C. J., and X. C. The interpretation was developed by L. K., H. D., J. Z. and Y. Z. The manuscript was prepared by H. D., J. Z. and M., T., together with help of all other co-authors. All authors discussed the results and contributed to the manuscript.

100408(R) (2011)

25. Zhang, X. C. *et al*. Skyrmion-skyrmion and skyrmion-edge repulsions in skyrmion-based memory. *Sci. Rep.* **5**, 7643 (2014)

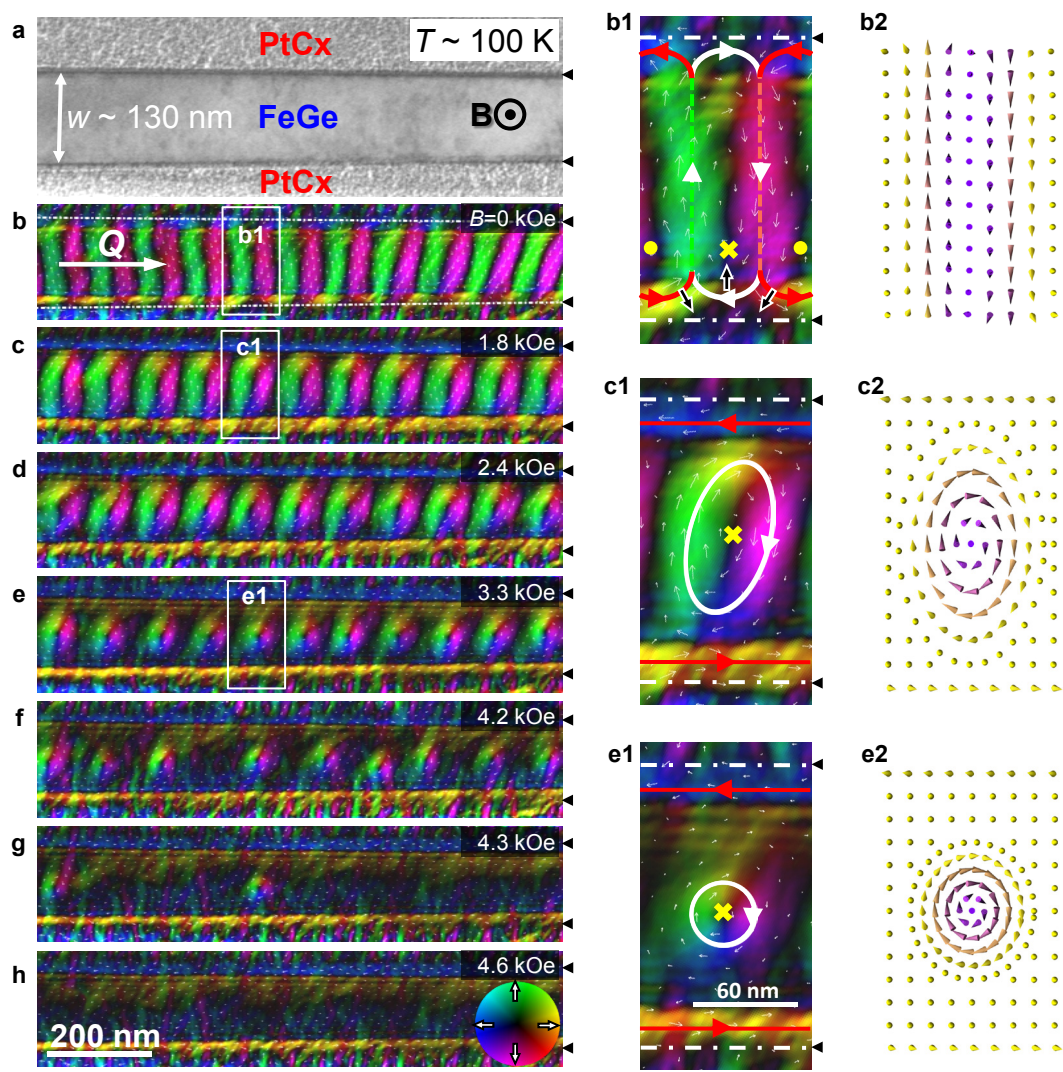
20

**Figure 1 Variations of spin texture with magnetic field in a 130 nm FeGe nanostripe at $T$ = 100 K. a,** TEM image of the FeGe nanostripe surrounded by an amorphous Pt layer. The magnetic field is applied normal to the stripe plane. **b-h**, Magnetic-field dependence of the spin texture, represented by the lateral magnetization distribution as obtained by transport-of-intensity equation (TIE) analysis of three Lorentz TEM images with the defocus values of -144 μm, 0, +144 μm. The color wheel represents the magnetization direction at every point. "**Q**" is the wave vector. **b1, c1 and e1**, Enlarged regions marked by the white boxes in corresponding panels (**b**) to (**e**). The small white arrows represent the in-plane magnetization direction at each point. "●" and "×" stand for the upward and downward directions of spins, respectively. At $B \sim 0$ Oe, spin helices are terminated at the edge and form two types of "half-disk" domains by distinguishing the curling direction of in-plane magnetization around the edge (thick red lines, anti-clockwise; thick white lines, clockwise). The magnetic field-driven evolution of "half-disk" domains is illustrated by using small black arrows to point out the moving directions of the domains in the bottom of panel **b1**. For clarity, **b2, c2 and e2** depict the schematic spin arrangements shown in the corresponding panels **b1** to **e1**. The white dot lines in panel b stands for the FeGe/PtCx interfaces, which are marked by small black triangular in the magnetic images (**b-h**) and (**b1** to **e1**).

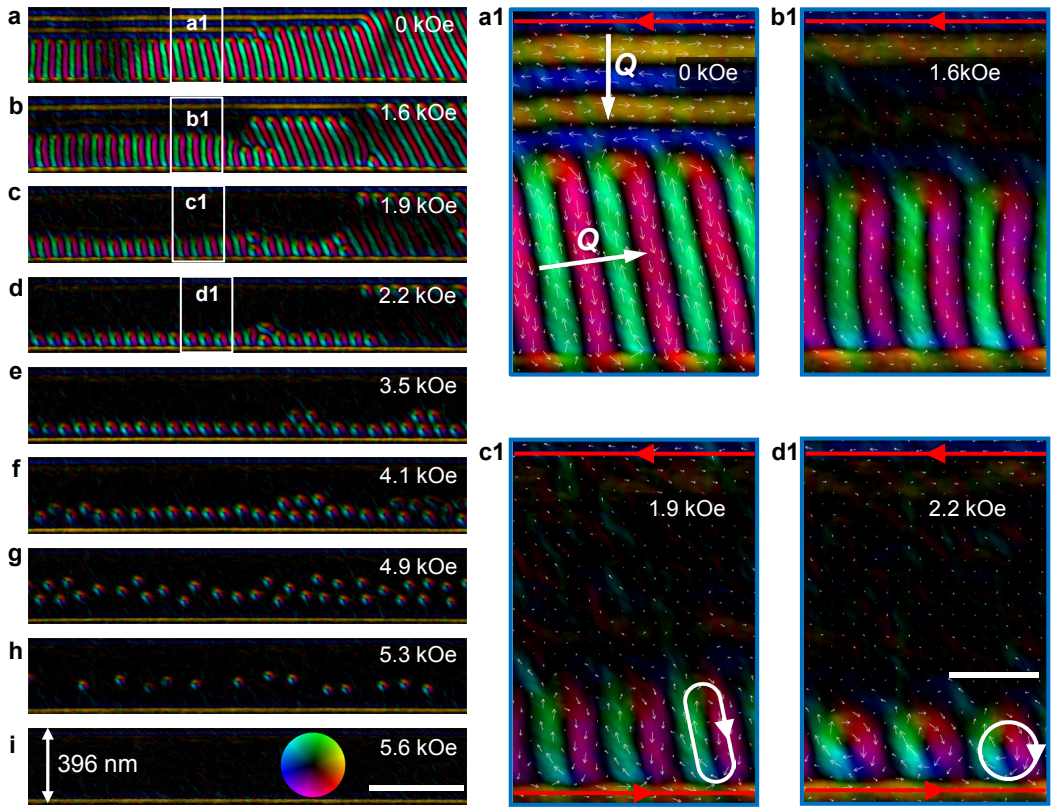

**Figure 2 Variations of spin texture with magnetic field in a 396 nm FeGe nanostripe at $T$ = 100 K. (a – d)**, The different behavior of spin helices with wave vectors **Q** ∥ edge and **Q** ⊥ edge in a magnetic field. (**a1- d1**), The corresponding enlarged region marked by the white boxes in the corresponding panels (**a**) to (**d**). Under the applied magnetic field, the spin helix with almost **Q** ∥ edge transfers into skyrmions, while those with almost **Q** ⊥ edge disappear directly without leaving skyrmions. (**e-g**), The collective movement of a skyrmion chain into the center of the stripe with increasing magnetic field. (**h-i**), The transition of isolated skyrmions to ferromagnetic spin textures with uniform edge state. The red arrows indicate the orientation of the magnetization around the edge. In order to show magnetic

structure in FeGe nanostripe more clearly, all the images are tailored to remove the PtCx part. The defocus values used for Lorentz TEM imaging are -192 μm, 0, +192 μm.

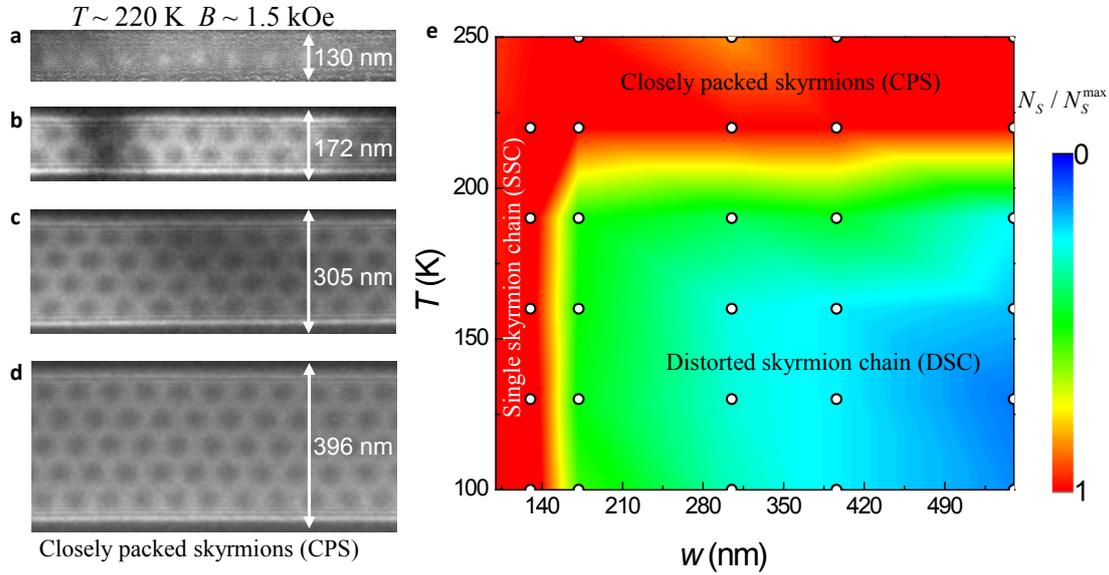

**Figure 3 Sample width dependence of the skyrmion arrangement in the temperature-width diagram. (a-d),** The width dependence of skyrmion arrangements obtained at an elevated temperature, $T = 220$ K as a magnetic field was applied. The image is acquired under over-focused condition with the defocus value 288 μm. The dark or bright circles represent the skyrmions. The change of dark or bright is due to the reversal of the crystal chirality. (**e**), Sample width dependence of the skyrmion phase diagram in the plane of temperature and width. The open dots are data points from the Lorentz TEM measurement. From these data, a colored map is constructed to show the normalized skyrmion density, defined as $N_S/N_S^{max}$, with $N_S$ the actual number of skyrmions at each temperature, and $N_S^{max}$ the maximum number of skyrmions that can be accommodated in the nanostripe.

# Supplementary Information

**Supplementary Note 1. Lorentz TEM and magnetic transport-of-intensity equation (TIE) analyses.**

The principles of Lorentz TEM may be understood classically in terms of the basic interaction between the electron beam and the magnetic fields within and around the magnetic specimen[1]. The most commonly used technique for revealing the domain structures is the Fresnel (or defocus) method[2-3]. In the method, the objective lens is defocused so that an out-of-focus image of the specimen is formed. The schematic ray diagram in the Fresnel model of the TEM is shown in the supplementary Figure 1. For the purpose of illustration, a simple specimen comprising two domains separated by 180º domain walls is assumed.

When the parallel electron beam passes through the area around the domain wall in the thin sample, Lorentz force, defined as $\mathbf{F} = e(\mathbf{v} \times \mathbf{B})$ with the electron velocity $\mathbf{v}$ and the magnetic induction $\mathbf{B}$, will lead to the deflections of the electrons. Following the right-hand rule, the electron beams, irradiated on the left and right domains with opposite in-plane magnetization orientations, are deflected in left and right directions, respectively. When imaging the domains in in-focused conditions, these deflecting electrons are focused in the final image plane so that no magnetic contrast appears (supplementary Figure 1a). By contrast, when the Lorentz TEM is in the over-focused conditions, the electron deflection induces a low intensity contrast since the electrons are deflected away from the domain wall. This results in the appearance of dark contrast line in the domain wall region (supplementary Figure 1b). Similarly, a bright contrast line appears in under-focused conditions due to the increased electron density caused by overlapping (supplementary Figure 1c). In this sense, the inversion of the magnetic contrast in the domain wall is observed between the over- and under-focused images. This is the common feature in the Fresnel model images. It is worthy noticing that the out-of-plane magnetic components can not affect the electron propagation so that Lorentz TEM can't detect the out-of-plane magnetic components. In this study, we changed the objective lens current to control the magnetic fields applied to the specimen along the z-axis.

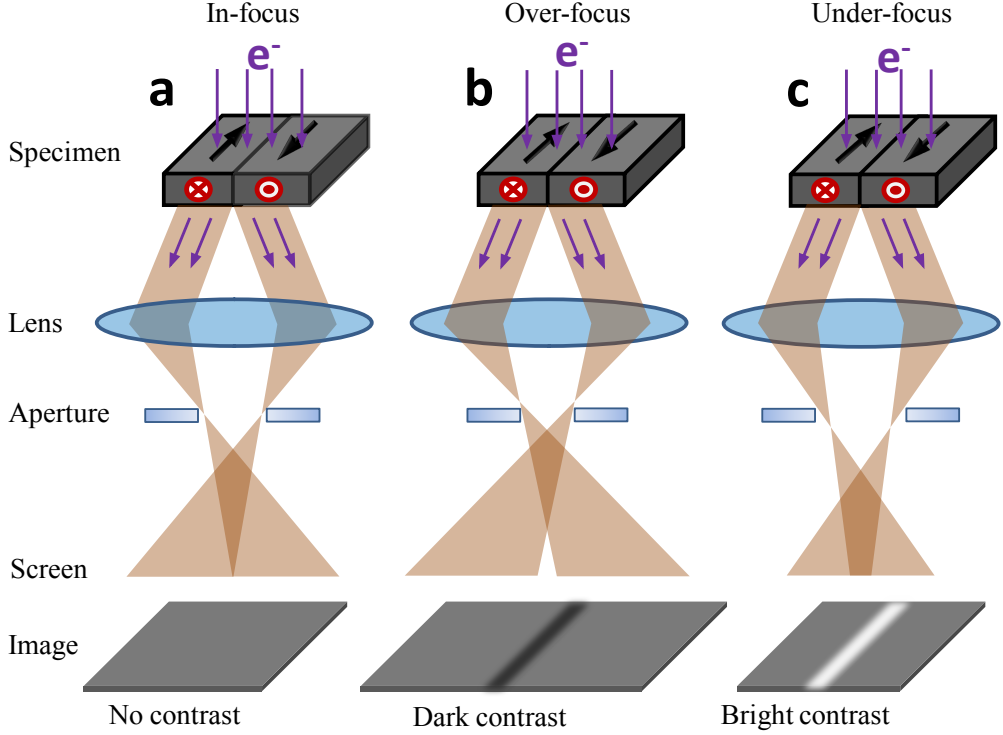

**Supplementary Figure 1 | The schematic ray diagram in a Fresnel image of a ferromagnetic specimen containing two 180º domain walls. (a),** In-focus conditions, the defected electron beams are focused in the final image plane so that no magnetic contrast appears. **(b)** and **(c),** Over- and under-defocus conditions, the defected electron beams lead to dark and bright contrasts, respectively.

From the above discussion that the magnetic contrasts in the domains wall depend on the defocused conditions it is possible to reconstruct and map the in-plane magnetic components distribution of the domain walls. For these purpose, a commercial software package QPt is used[2,3], where three Lorentz TEM images at different defocus value (under-, in-, and over-focus) were analyzed by using the transport-of-intensity (TIE) equation.

$$\frac{2\pi}{\lambda}\frac{\partial I(x,y)}{\partial z} = -\nabla_{xy} \cdot [I(x,y)\nabla_{xy}\phi(x,y)] \quad (1)$$

where $I(x,y)$ and $\phi(x,y)$ stand for the intensity and phase distributions of propagating wave distribution, respectively. $\lambda$ is the electron wavelength. On the other hand, according to Maxwell-Ampére equations, $\phi(x,y)$ and magnetization **m** has a relationship

$$\mathbf{m} \times \mathbf{n} = -\frac{\hbar}{et}\phi(x,y) \quad (2)$$

where $e$, $\hbar$ and $t$ are the electron charge, the reduced Planck constant and the thickness of the sample, respectively. **n** is the unit vector parallel to the beam direction. The in-plane magnetic components can be obtained as the phase shift $\phi(x,y)$ is known, The intensity gradient $\partial I/\partial z$ can be approximately expressed as $\Delta I/\Delta z$, considering that defocus step $\Delta z$ is far less than focal length. In this study, a typical process to obtain the magnetic components of skyrmion phase is shown in supplementary Figure 2.

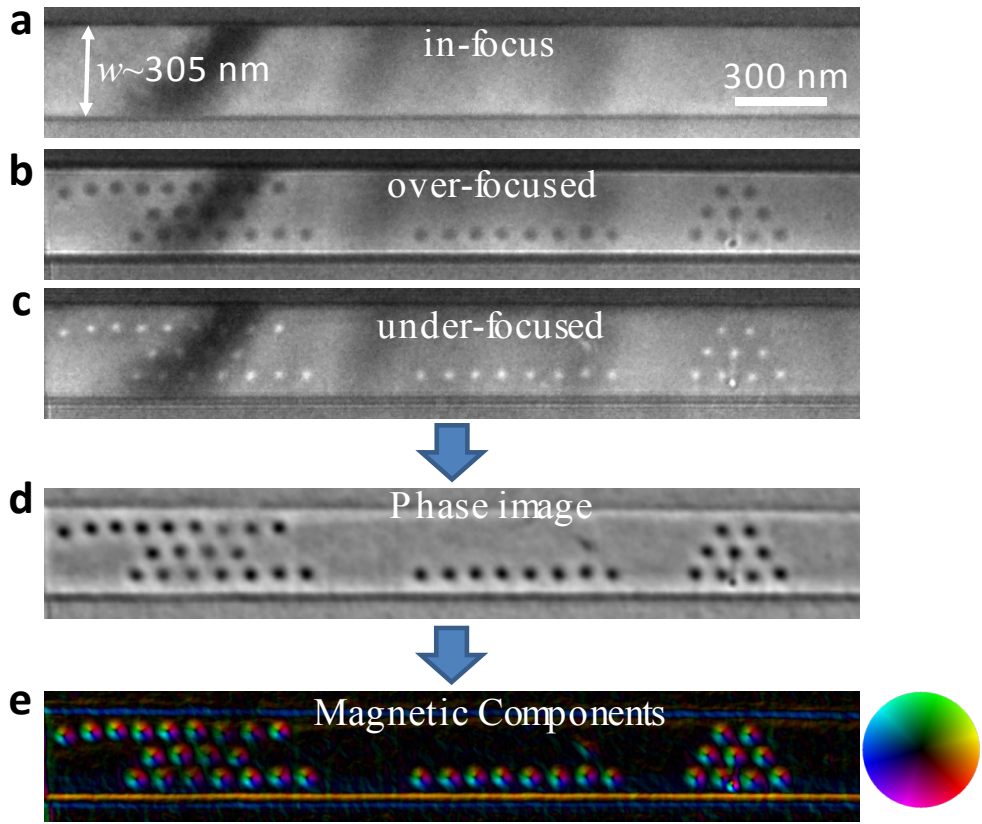

**Supplementary Figure 2 | Schematic diagram of magnetic TIE analysis in the skyrmion state.** Three images in the same region under different defocus conditions (**a**, in-focused; **b**, over-focused; **c**, under-focused; $\Delta z = 196\ um$) are acquired by using Lorentz TEM. (**d**), The phase image was created from the three Lorentz micrographs applying Eq. (1). (**e**), The in-plane magnetization distribution map was

generated from phase image by applying Eq. (2). The sample is named as NS3 with its detailed description in supplementary Table 1 and supplementary Figures 4 and 5. The color wheel represents the magnetization direction at every point. These color wheel is used in the whole text including the main part and supplementary information.

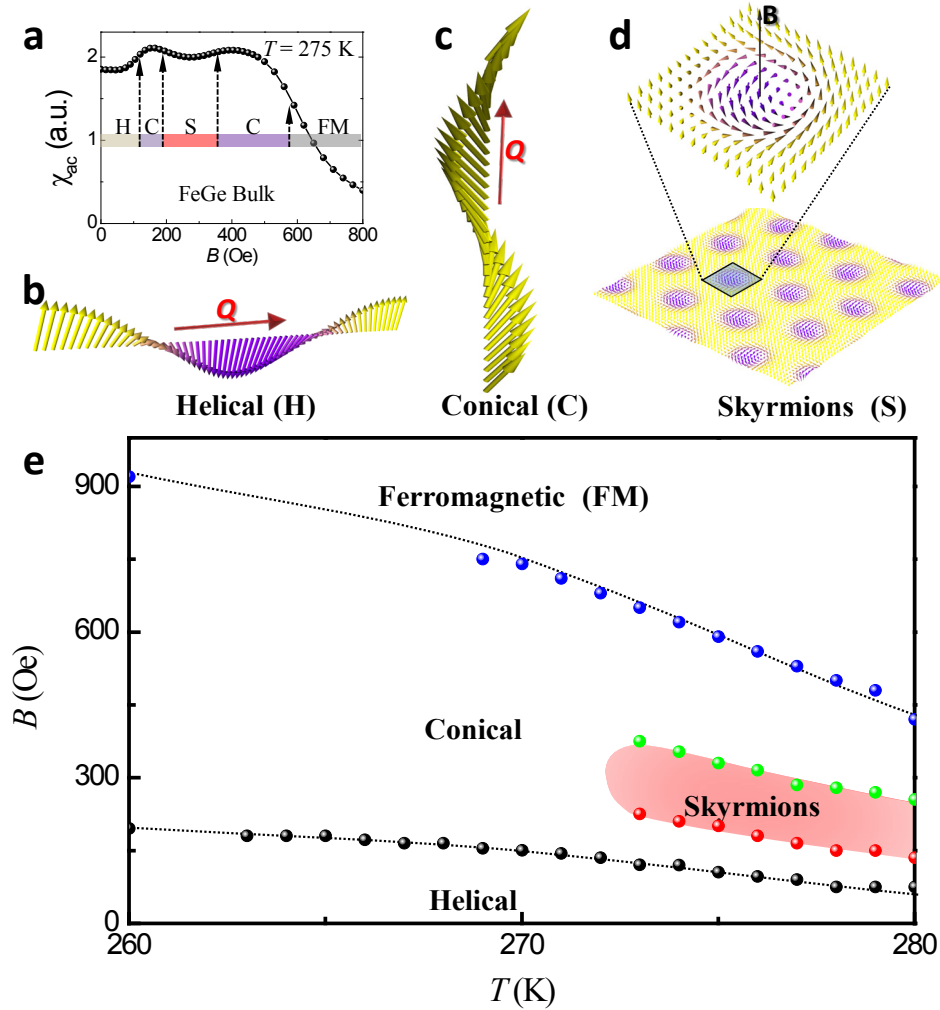

**Supplementary Figure 3 | Magnetic phase diagram of bulk FeGe inferred from the ac susceptibility. (a),** Typical isothermal ac susceptibility data at $T \sim 275$ K. Subtle changes in $\chi_{ac}$ ($B$) below 800 Oe indicate rich magnetic phases, which are identified by comparing the $\chi_{ac}$ ($B$) curve with their characteristic forms in helical magnets[5,6]. **(b), (c),** and **(d),** schematically, represent the spin configurations of helical, conical and skyrmion phases, respectively, which develop below $T_c \sim 280$ K with the increase of magnetic field. Clearly, the helical order unpins under application of a

high magnetic field to form the conical phase with its wave vector along the direction of magnetic field, and then condenses into skyrmion crystal with a hexagonal arrangement. (**e**), Phase diagram of magnetic structure in bulk FeGe inferred from the ac susceptibility. The ac susceptibility was measured on a Quantum Design physical property measurement system with the excitation amplitude of 3.5 Oe and excitation frequency of 333 Hz.

**Supplementary Note 2. Fabrication and characteristic of FeGe nanostripes.**

We fabricated the FeGe nanostripes by a top-down method from the bulk sample. Supplementary Figure 4 schematically illustrates the total process, which are depicted as follows

**Step 1**: Following the standard TEM specimen preparation procedure[6], a homogeneous FeGe thin narrow membrane with the desired thickness is carved on the surface of a FeGe bulk by focused ion beam (FIB) milling. The amorphous surface layer induced by the high energy FIB gallium beam is then reduced to about ~2 nm by polishing the surface with the low energy Ga beam.

**Step 2**: Using the Gas Injection System (GIS), an amorphous PtCx film was deposited on either side of the membrane by means of e-beam evaporation. These PtCx layers are nonmagnetic and only used to reduce the Fresnel fringes at the edges, as discussed later. Hence, they have no effect on the magnetic properties of the FeGe nanostripe.

**Step 3**: Using the FIB, the sandwich structure PtCx/FeGe/PtCx is carved to U-shape to prepare the lift-out.

**Step 4**: Using the Omniprobe 200+Micromanipulator, the membrane with two PtCx coating layers is released from the bulk and then transferred onto a clean silicon substrate.

**Step 5**: Using the standard TEM specimen preparation method again, a PtCx/FeGe/PtCx sandwich structure is fabricated by FIB milling.

**Step 6**: Using the FIB, the desired sample is carved to U-shape to prepare the lift-out.

**Step 7**: Using the Micromanipulator, the Pt/FeGe/Pt sandwich structure was attached to a TEM Cu chip for the final Lorentz observations.

By adjusting the parameters in the sample preparation process, a variety of TEM specimens are achieved. Parameters of FeGe nanostripes for Lorentz TEM imaging are shown in supplementary Table 1

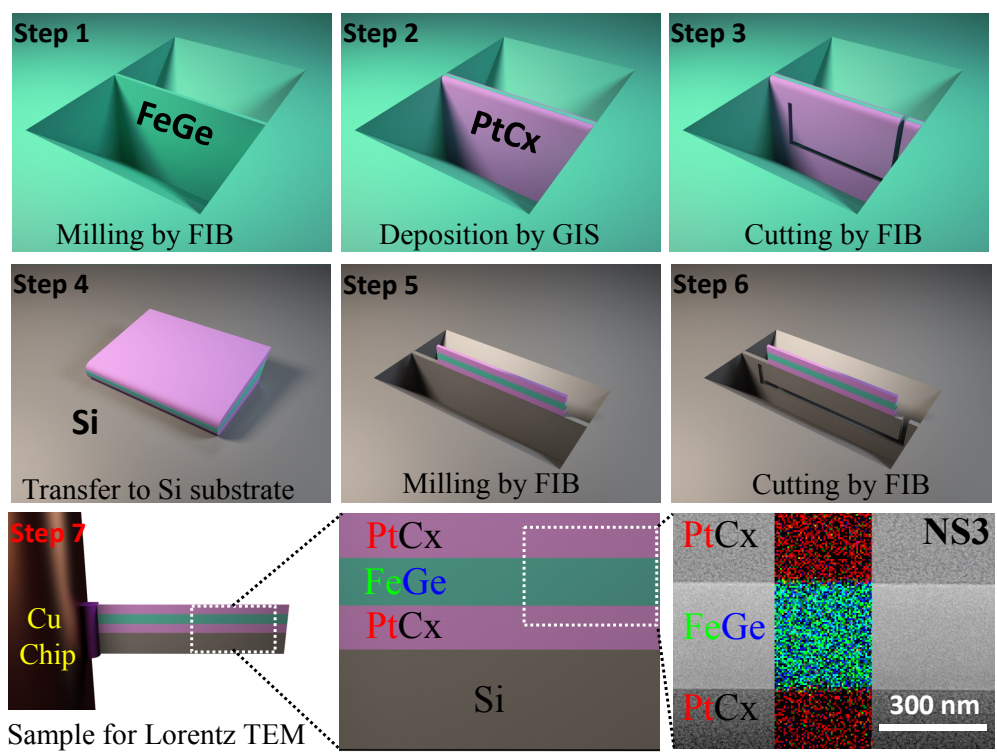

**Supplementary Figure 4 | Schematic procedure for fabricating the nanostripes by using focused ion beam and scanning elctron microscopy (FIB-SEM) duel beam system (Helios Nanolab. 600i, FEI) equipped with a Gas Injection System (GIS), and Omniprobe 200+ micro-manipulator.** The whole process is schematically depicted from step 1 to 7. A typical nanostripe is shown in the final panel. The compositions are measured by energy dispersive spectroscopy (EDS, Oxford X-MaxN 80T) under scanning TEM (STEM) model with the operated voltage 200 kV. The compositional maps for Fe, Ge, and Pt with color superposition (Fe blue; Ge Green; Pt; red) show the clear FeGe/PtCx interface. The coating layer is the nanocrystal Pt with a typical size of 3-5 nm embedded into amorphous carbon matrix[7]. Since carbon is light elements and the operated voltage in STEM mapping is high, the carbon is hardly resolved under this condition.

**Supplementary Table 1 | Parameters of FeGe nanostripes for Lorentz TEM imaging**

| Samples Name | Width(nm) ($w$) | Crystal plane | Thickness (nm) ($t$) |
|---|---|---|---|
| NS1 | 130 | [-123] deviation 5° | 65 |
| NS2 | 172 | [111] deviation 7° | 85 |
| NS3 | 305 | [111] deviation 7° | 82 |
| NS4 | 396 | [111] deviation 7° | 91 |
| NS5 | 550 | [112] deviation 15° | 92 |
| NS6 | 1017 | Polycrystalline | 90 |

**Supplementary Note 3. Comparison of Fresnel fringes at FeGe/PtC$_x$ and FeGe/Vacuum interfaces.**

In transmission electron microscopy, Fresnel contrast occurs if the observed region in which the projected electromagnetic potentials – either mean inner potential or magnetic potential or thickness - changes abruptly and is imaged under out of focus conditions[8]. The Fresnel fringe contrast is often seen at the edge of an object imaged under out-of-focus. For magnetic characterization, the Fresnel imaging technique, as discussed in supplementary note 1, enable direct observation of magnetic structure e.g. domain wall, provided that the thickness variation or projected electrostatic potentials are negligible small as compared with the magnetic induction contributions[9]. However, at specimen edges, variation in Fresnel fringes due to the abrupt change in thickness overshadows the contrast change due to magnetic potential. This makes the analysis of magnetic information at the edge extremely difficult. Previous Lorentz TEM investigation on FeGe thin plates has illustrated the artificial magnetic contrast due to the Fresnel fringes extends above ~100 nm[3]. Concerning the helical period of FeGe is ~ 70 nm. This is sufficient to completely eradicate or severely distort the real domain structure of interest edge.

In this study, the use of amorphous PtCx adjacent to the edge of FeGe greatly reduces the effect of Fresnel fringe at the specimen edges due to the much-reduced variation in thickness, as opposed to the case of vacuum-edge. A direct comparison of Fresnel fringes at FeGe/PtCx and FeGe/vacuum interfaces is illustrated in supplementary Figure 5. For this FeGe nanostripe, the Pt layers were firstly coated on two side of the stripe, and the right PtCx layer is fall off by one chance. This thus enables us to directly compare the influences of the Fresnel fringes at the two FeGe/PtCx and FeGe/vacuum interfaces. Supplementary Figure 5a,b,c show the TEM images of the two interface under in-, over- and under-focused conditions, respectively. The yielded magnetic components are shown in supplementary Figure 5d, where the significantly reduced Fresnel effect at the FeGe/PtCx interface is clearly observed. It is therefore possible to investigate the magnetic helix and skyrmion states in geometrically confined nanostructure by real-space Lorentz microscopy observation.

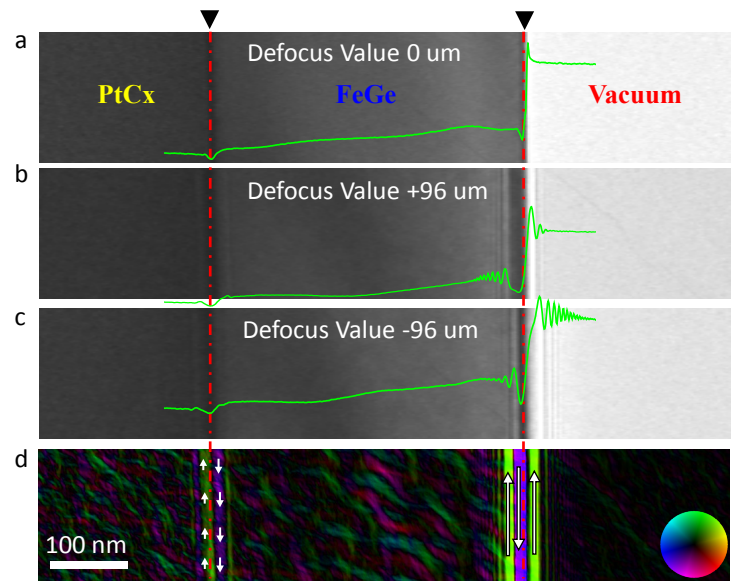

**Supplementary Figure 5 | Comparison of Fresnel fringes at Pt/FeGe and FeGe/vacuum interfaces.** (**a**), Under-focused case with the defocus value 96 μm; (**b**), in-focused case; (**c**), over-focused case with the same defocus value 96 μm. It is clearly shown that the range of Fresnel fringes extends above ~100 nm at the FeGe/vacuum interface in the under-focused or over-focused conditions, while it is below ~50 nm at the Pt/FeGe interface (**a** and **c**). Moreover, compared with the FeGe/vacuum interface, the strength of Fresnel fringes at the FeGe/PtCx interface is also significantly weakened. This advance enables the Lorentz TEM to directly image the edge state with less influence by the Fresnel fringes. The red dot lines indicate the position of the FeGe/PtCx and FeGe/Vacuum interface. For clarity, the corresponding strength of Lorentz TEM images under different focused conditions are plotted as the green blue lines. The white arrows represent the direction of lateral magnetization.

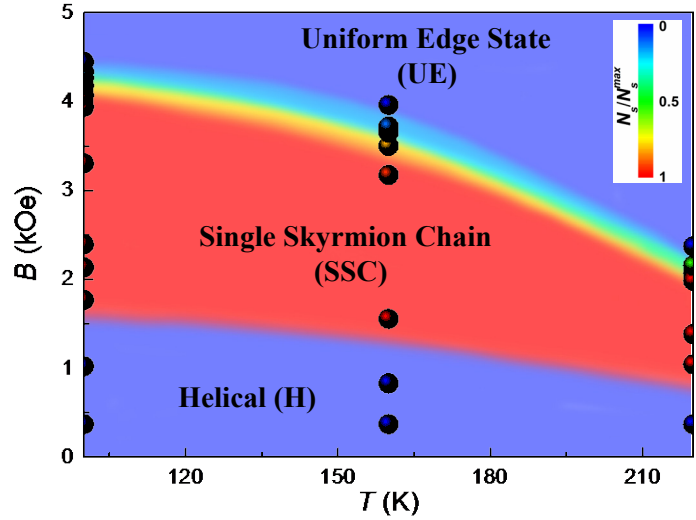

**Supplementary Figure 6 | Magnetic phase diagram of the 130 nm nanostripe in the *T-B* plane.** SSC represents the single skyrmion chain. The colored dots show the experimental points obtained by mapping of the normalized skyrmion density, from which the region of hosting SSC are marked in red for reference.

**Supplementary Note 4. Analysis of the magnetic contrast around the edge.**

Here we presented the edge state in 305 nm nanostripe, as shown in supplementary Figure 9, an enlarged scale to permit closer inspection of the detailed spins arrangements around the edge. Supplementary Figure 6a shows the spin texture at zero magnetic field, where a periodically modulated "half-disk" domains, marked by white (clockwise) and red (counter-clockwise) arrows, is clearly observed. This observation is the common feature in the nanostripes only if **Q** ∥ edge. It is noteworthy that the FeGe/PtCx interface (supplementary figure 5) still leads to weak artificial magnetic contrasts though its strength and extended regions is significantly reduced. As a result, the observed magnet components around the nanostripe edge are the superposition of real and the weak artificial ones. To analyze the real magnetic structures, we perform the Lorentz TEM measurement on the nanostripe in the same region at $T \sim 300$ K. Since the temperature $\sim 300$ K is far above the Curie temperature $T_c \sim 280$ K of the FeGe sample[5], the magnetic contrast of the edge comes solely from the edge Fresnel fringes. Supplementary Figure 6b shows the yielded artificial magnetic arrangements around the edge of the nanostripe, where the orientation of the moments is along the right direction (marked by red arrows). This orientation is consistent with the counter-clockwise "half-disk" domains. This means the real counter-clockwise and clockwise "half domains" are strengthened and weakened, respectively, by the artificial magnetic contrast. A simple assumption is that these half domains are degenerate since no extra effects are able to break the degenerate at zero magnetic field. This assumption is supported by the model calculations[10]. Accordingly, we proposed the real magnetic structure with degenerate "half domains" around the edges (supplementary Figure 6c).

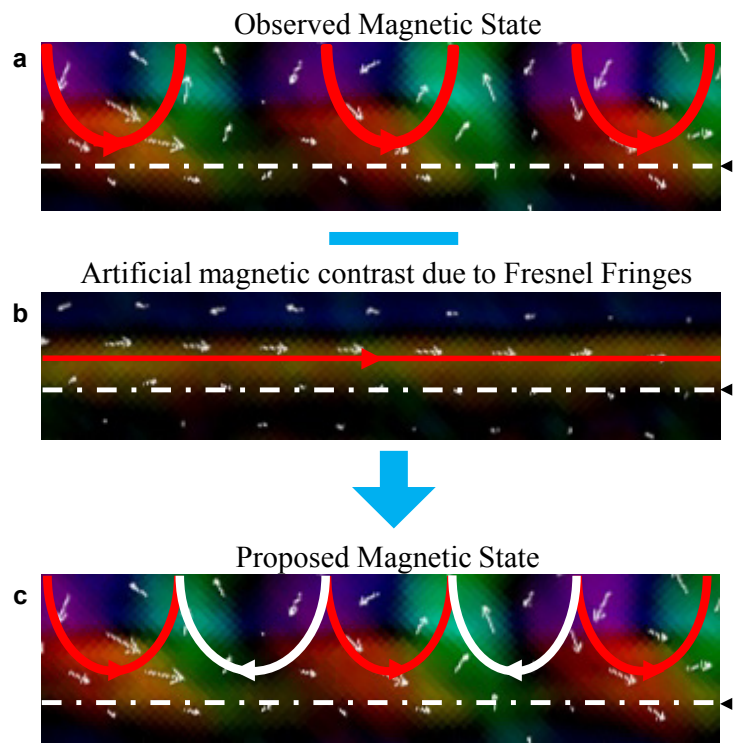

**Supplementary Figure 7 | Magnetic structure around the edge of a 305 nm nanostripe at zero magnetic field. (a)**, the observed spins arrangements. The red and white arrows stand for the counter-clockwise and clockwise "half domains", respectively. **(b)**, the artificial spins arrangements due to the Fresnel fringes. The red arrows stand for the direction of the artificial magnetic moments. **(b)**, The proposed spins arrangements with the degenerate "half domains". The white dot lines indicate the position of the FeGe/PtCx interface.

When the magnetic field is applied, the degenerate edge state is broken by forming the skyrmions in the interior (Figure 1 in the main text). In this case, a uniform edge state appears. According to the theoretical prediction that a magnetic field will weaken the edge lateral magnetization, but can't fully polarize the edge spins even at the highest magnetic field due to the boundary conditions[11-14]. To test this predication, we plotted the lateral magnetization of the uniform edge state as a function of the external magnetic field. By roughly measuring the arrows length around the edge, we clearly observed the lateral magnetization of the uniform edge state decreases with increasing magnetic field (supplementary Figure 7e), which is accordance with the theory. On the other hand, the above-mentioned artificial magnetic contrast contributes to the observed magnetic contrast. For comparison, the artificial magnetization is also plotted by the open star (Supplementary Figure 6b). It shows the real in-plane magnetic components can't be well distinguished from the artificial magnetic contrast at high magnetic field.

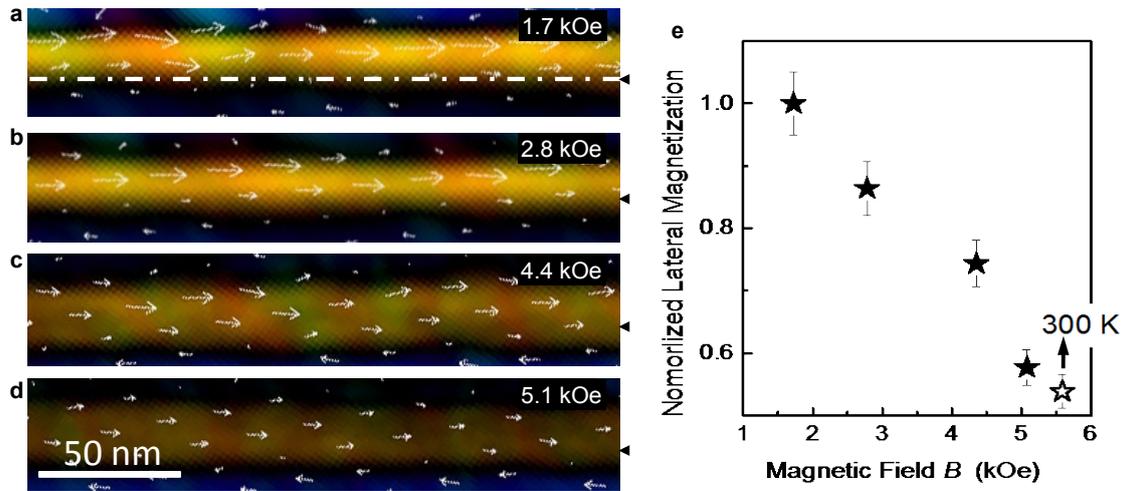

**Supplementary Figure 8 | Magnetic field dependence of reduced lateral magnetization at the edge of a 305 nm nanostripe. (a-d),** Spin arrangements at the edge with varied magnetic field obtained at $T \sim 100$ K. **(e),** The reduced magnetization as a function of the magnetic field. For comparison, the artificial magnetization is also plotted by the open star

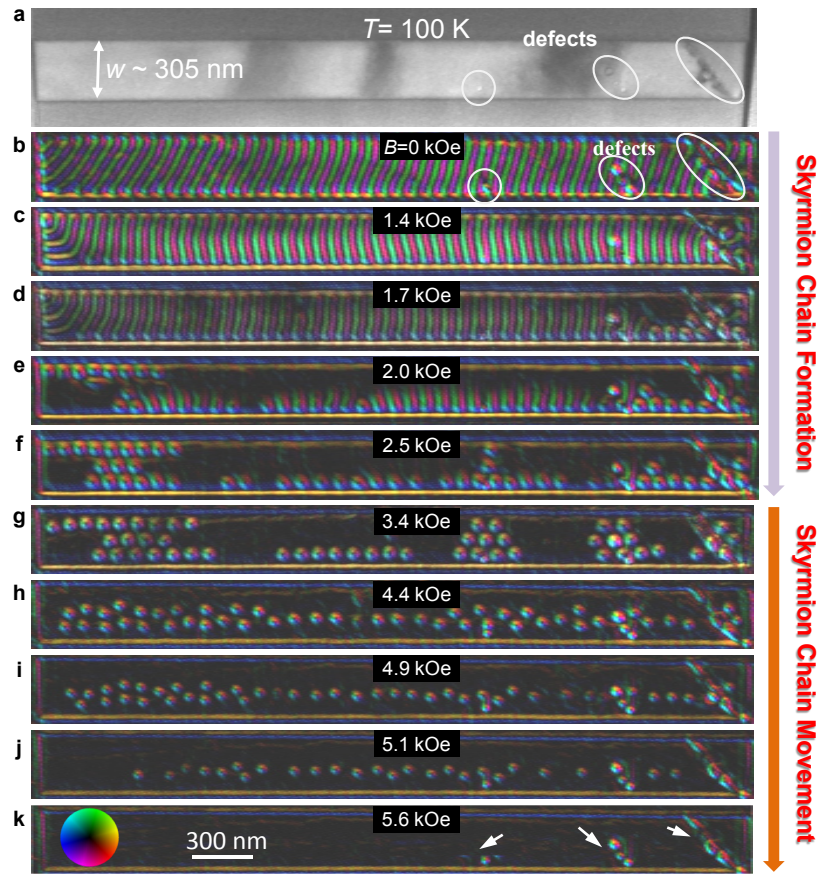

**Supplementary Figure 9 | The magnetization processes of a 305 nm FeGe nanostripe. a,** TEM image of the nanostripe. (**b-k**), Magnetic-field dependence of the spin textures obtained at 100 K. The white circles highlight the defects in the sample. At $B \sim 0$ Oe, the spins show a helical ground state with **Q** ∥ edge, the same to 130 nm stripe. When the magnetic field is switched on, the nanostripe shows essentially similar dynamics of magnetization with 130 nm or 396 nm nanostripes presented in the main text. Specifically, the self-organized skyrmion chains sit at the edges at low magnetic field, then decouple from the distorted edge state and move to the inner region of the nanostripe with increasing magnetic field. Defects including corner, dot or grain boundaries are also beneficial for the creation of skyrmions[2,3]. At low magnetic fields, skyrmions gathered together around the defects to form clusters (**f** and **g**), but the total number of skyrmions in the stripe keep unchanged.

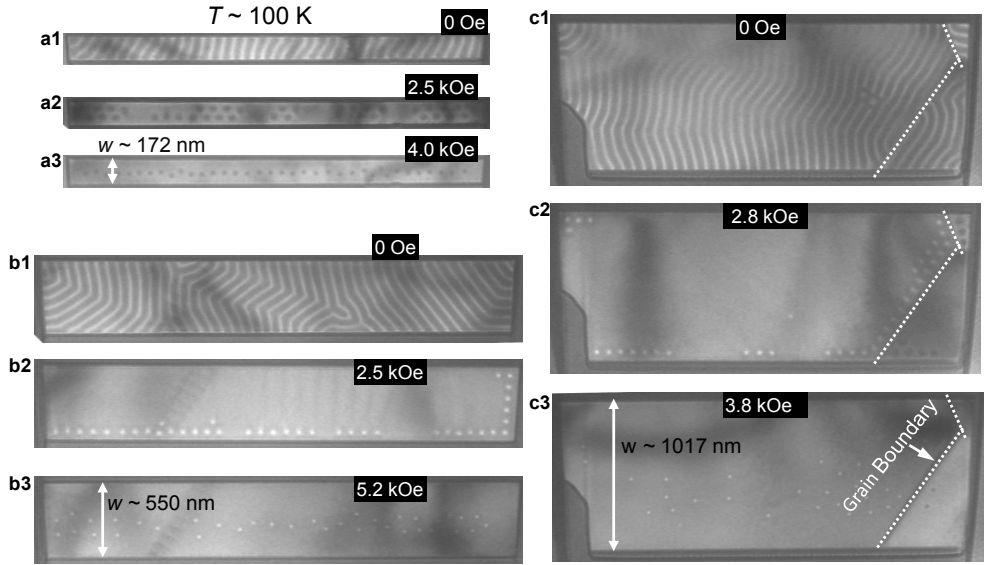

**Supplementary Figure 10 | The three typical magnetic structures (helical state, skyrmions around the edge, distorted skyrmion chain) of FeGe nanostripes with three widths at $T$~ 100 K under the applied magnetic field. (a),** $w$ ~ 172 nm. **(b),** $w$ ~ 550 nm. **(c),** ~ 1017 nm. All samples follow similar magnetization dynamics as shown in Fig. 2 that the helical states with distorted edge spins evolve into the skyrmions, which assemble in the form of chain around the edge at low field and move collectively into the center of the nanostripes at high field. This observation indicates that the mechanism of the edge-mediated skyrmion chain is independent of the sample width. Notice that the grain boundaries may also play a similar role to the edge in creating skyrmions at low temperatures, as shown in the 1017 nm nanowire[9]. It is worthwhile mentioning that the skyrmion chain or helical state is able to move or oscillates under the action of the magnetic field, especially in wide samples. The high mobility of skyrmions leads to deviations of the skyrmion to its initial positions. But, there is nearly a one-to-one relationship between the number of created skyrmions and the completed edge distortion. In particular, for the polycrystalline 1017 nm sample, two small regions divided by grain boundaries meet this one-to-one correspondence too. In the wider samples, the skyrmion chain preferentially forms along the bottom edge of the nanostripe, which is probably due to the slightly inhomogeneous thickness of the nanostripes. In a narrow stripe $w$ ~ 172 nm, skyrmions formed on both sides. We have also fabricated samples with larger size, but the increased defects such as the grain boundaries and the inhomogeneous thickness make the situation more complicated. A detailed discussion of these results is beyond the scope of this paper. The images are acquired under over-focused conditions with the defocus value 192

µm. The dot lines indicate the grain boundary. The skyrmions are displayed as black or white circle dots. The change of magnetic contrast comes from the reversal of the chirality of the samples.

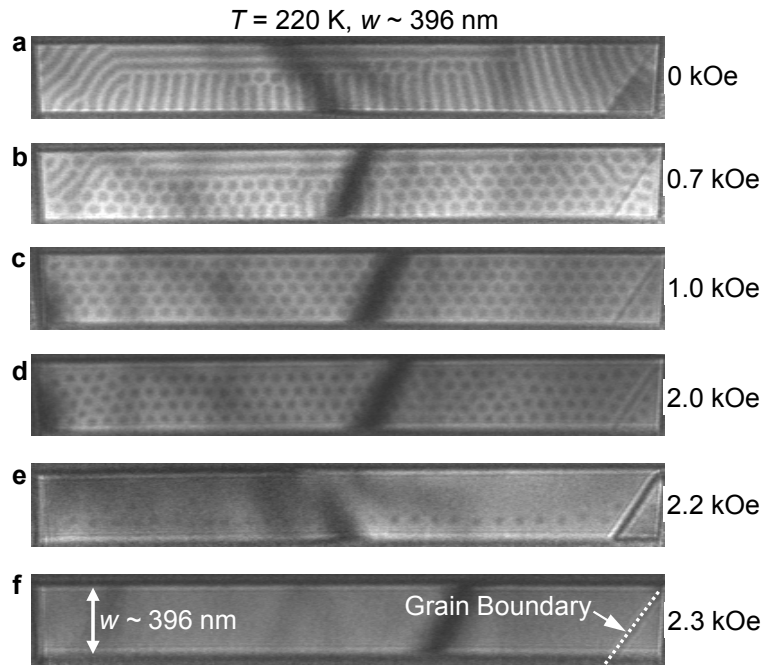

**Supplementary Figure 11 | The magnetic field dependence of spin textures in the 396 nm FeGe nanostripe at high temperature $T \sim 220$ K.** (**a**), The helical state; (**b**), mixture state of skyrmion lattice and helical phase; (**c**) and (**d**), a packed skyrmion lattice; (**e**), isolated skyrmions; (**f**), a field-polarized ferromagnetic state surrounded by edge vortex state. The images are acquired under over-focused conditions with the defocus value 288 µm. The dot lines indicate the grain boundary.

**Supplementary References**